# MOBILE AGENT BASED SOLUTIONS FOR KNOWLEDGE ASSESSMENT IN eLEARNING ENVIRONMENTS

Mihaela Dinsoreanu
Cristian Godja
Claudiu Anghel
Computer Science Department
Technical University of Cluj-Napoca
Baritiu 26-29, RO-3400
Cluj-Napoca, Romania
E-mail:mihaela.dinsoreanu@cs.utcluj.ro

Ioan Salomie
Tom Coffey
Department of Electronic and Computer Engineering
University of Limerick, Ireland
E-mail: {Ioan.Salomie, Tom.Coffey}@ul.ie

## ABSTRACT

E-learning is nowadays one of the most interesting of the "e-" domains available through the Internet. The main problem to create a Web-based, virtual environment is to model the traditional domain and to implement the model using the most suitable technologies. We analyzed the distance learning domain and investigated the possibility to implement some e-learning services using mobile agent technologies.
This paper presents a model of the Student Assessment Service (SAS) and an agent-based framework developed to be used for implementing specific applications.
A specific Student Assessment application that relies on the framework was developed.

## INTRODUCTION

Almost every domain we know has nowadays its "e-" Internet-based counterpart. We talk about e-commerce, e-banking, e-learning etc. Each "e-"domain emulates the traditional one in a new, virtual, Web-based environment. The major problems of creating the virtual environment involve traditional domain modeling and implementing the model using the most suitable technologies.
Our research is concerned with creating Web-based services for Virtual Learning Environments (VLE). This involves a complete analysis of the learning domain. The outcome of the analysis is the identification of the main concepts and relationships and building a conceptual model of the domain. On the other hand, the most appropriate technologies for implementing the model have to be analyzed and decided upon.
In this paper we focus on one aspect related to VLE, the Student Assessment. One of the most important educational components is the assessment of the student's acquired knowledge. There are several issues related to assessment that should be considered: communication issues, security issues, evaluation types, student answer analysis and grading.
This paper is structured as follows: an analysis of the Student Assessment domain is presented in Section 2, considering the most important concepts, constraints etc and building the conceptual model of the domain.
Section 3 presents a possible solution based on the Mobile Agents Technology. The concepts in the Application Domain are mapped in the Solution Domain, providing therefore a computational model. The computational model was further developed as a framework that can be used for implementing specific applications.
Section 4 presents a specific Student Assessment application that was built using the framework mentioned above.
We end with a discussion of some conclusions and possible developments in Section 5.

## VLE

VLEs have to provide all the necessary resources for overcoming time and space limitations existent in traditional f2f environments. Students and Instructors involved in a VLE can be located world-wide, they don't have to synchronize their communication, and their number is not limited.
Therefore, the services provided by VLEs should be designed considering issues like accessibility, scalability, security, communication etc. In our paper we will focus on one of the services of a VLE: the Assessment Service (AS).

### Student Assessment

AS provides the means of evaluating the students' acquired knowledge. It also provides the means for a student to get valuable feedback regarding his progress. AS is a highly dynamic component of the VLE, involving both synchronous and asynchronous communication between students and instructor. In order to build a model of AS, we analyzed the assessment process, different possible scenarios, and different assessment types. Based on the analysis we identified the main concepts involved and the relationships between them.

### Main Concepts Identification

Analyzing the main concepts involved in student evaluation we identified the following:
- Learning entity (the Student)
- Teaching authority (the Instructor)
- Assessment type (Compulsory Examination, Self-Assessment)

- Test
- Question Type
- Question
- Correct Answer
- Assessment procedure (as an Evaluation Engine)

The relationships between the concepts are depicted in a simplified manner in Figure 1.

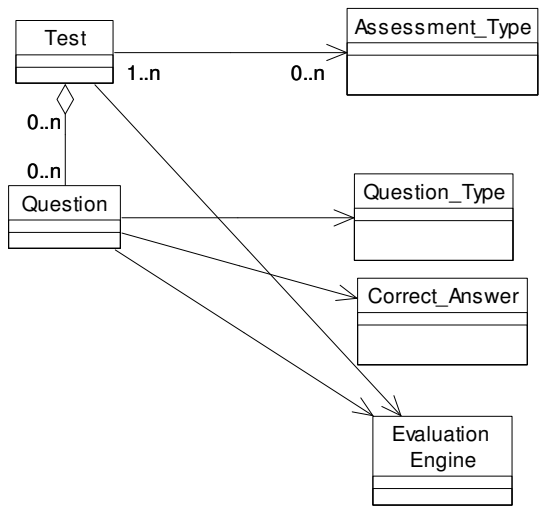

Figure 1: Assessment Concepts and Relationships

The Teaching Authority provides the tests. A test may belong to different assessments (Compulsory examinations, self-assessments, etc) and contains a set of questions. A question is associated to a question type, to one or more correct answers and also to an assessment procedure (implemented as an evaluation engine).
The evaluation engine provides the knowledge for evaluating the Student's answer against the correct answer(s) associated to a question. Student answers can be in a limited range, long natural language essays can not be analyzed.
The Student is able to access available tests, to run a test and provide his answers. The student should also receive feedback regarding his performance (the grade, the correct answers etc).

**Main Functional Tasks**

The next step in the analysis of the system is the representation of the functional requirements of the system. Considering the main external actors that interact with the system, the Student and the Instructor, we modeled the functional requirements in an UML-based manner as use-cases associated to the external actors.
The main functionalities of the system provided for the Student are: Visualization of Tests, Start a new Self-Assessment Test, View Test Results.
The functionalities provided for the Instructor are mainly: View Tests, Add new Test, Modify existing Test, Delete existing Test, View taken Tests, Schedule Tests
Other administration related functionalities are also considered

**Non-functional Requirements**

Besides the main concepts and functionalities described above, when modeling the AS service, some additional constraints have to be considered: independence of the VLE, independence of the implementation technology, scalability and accessibility. Another constraint is related to question/answer types. As mentioned above we did not consider Student answers as essays, therefore AS is more suitable for technical disciplines where the correct answers are in a limited range.

**MOBILE AGENTS – AN EFFICIENT SOLUTION**

Since we are dealing with a highly distributed system and considering the constraints mentioned above, we investigated the possibility to provide a solution based on agent technology. Our goal was to design a multi-agent system that fulfills the functional requirements described respecting also the discussed constraints.
In our approach the multi-agent system is considered an organization of agents. The organization knowledge and capabilities are larger than the sum of knowledge and capabilities of the individual agents (Wooldridge et al. 2000), (Zambonelli et al. 2000), (Zambonelli et al. 2001).
In modeling organizations the following factors should be generally modeled at some level of detail (Weiss 1999):
- Agents comprising the organization
- The organization's design (structure)
- Tasks that should be carried out
- The environment the organization exists in
- Stressors acting on the organization and

We started our design by mapping the functional requirements represented as use-case diagrams to a set of tasks the system has to perform.
In order to build a complete task model we considered a top-down approach decomposing more general tasks to specific subtasks. Next, we identified the necessary agent roles to perform the tasks. Agent roles define the position of the agent in the organization.
The organizational design actually consists of a set of models, each addressing one facet of the organization:
- Environment Model
  The Environment model represents the available resources and also access protocols to resources.
- Interaction Model
  The Interaction model represents the communication structure between agents.
- Role Model
  The Role model is actually the authority structure in the organization. It links also tasks to roles.

**Task Decomposition**

Analyzing the use case diagrams that model the functional requirements of the system, we considered the following main tasks:

**Communication Tasks**
Communication is a key issue from both internal and external viewpoints. The organization obviously does not exist in isolation so it has to communicate to the exterior world. On the other hand we talk about an organization, so agents are supposed to communicate in order to achieve their goals. Therefore, we considered the two main communication types:
- Communication to external actors (Student, Instructor, VLE)
- Communication inside the system (modeled by Interaction Protocols)

To provide efficient communication to human external actors a **Personal Assistant Agent** was considered. The Personal Assistant (PA) is a stationary agent living on the client machine and providing the communication interface between the external actor (Student, Instructor) and the system.

**Coordination Tasks**

Besides communication, coordination of the organization is also a key issue. Coordination tasks involve: handling self-assessment requests, handling compulsory examinations, generating evaluation engines, performing evaluation etc.

The system was designed as a centralized coordinated system, the core of the coordination module being a **Server Agent.** The Server Agent (SA) is a stationary agent that lives on the AS machine and is responsible with handling self-assessment requests, examinations set by Instructors, generating corresponding evaluation engines etc.

For the evaluation itself we considered an **Evaluation Agent.** The Evaluation Agent (EA) is a mobile agent that migrates on the client (Student) machine and is able to perform the evaluation. EA is loaded with an Evaluation Engine containing the complete Test (questions, answer options, correct answer) and the assessment procedure.

The creation of the EA is also SA's responsibility.

We also considered other dependencies between tasks (Weiss 1999): pooled (results of one or more tasks jointly needed to perform another task), sequential (two or more subtasks should be performed in a specific sequence), reciprocal (two tasks depend jointly on each other) .

**Organizational Model**

As previously stated, modeling an organization involves several concepts comprised in different sub-models of the organization. Our approach models a closed organization (no alien agents are allowed), containing benevolent, cooperative agents. We considered the following sub-models as components of our organizational model.

**Environment Model**

The Environment Model represents the resources available to the agents and the associated access protocols to them. We consider as resources both data and knowledge storage structures and other components (objects, servers etc) that provide specific services to agents. The design of the agents is independent of any specific resources. The Environment Model is represented by several UML-based package and class diagrams.

**Role Model**

This model contains the agent roles in terms of their tasks, interactions and accessible resources. As mentioned above we identified three agent roles like depicted in Figure 2.

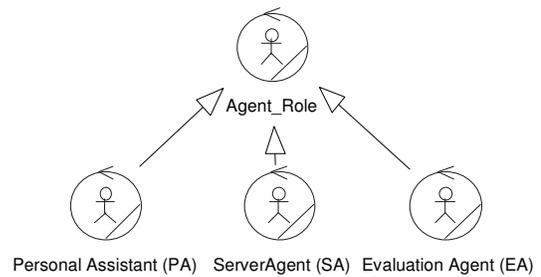

Figure 2: Agent Roles

Each role is associated to the set of tasks it's responsible for. We modeled the tasks as UML-type use-cases. In Figure 3 EA and its associated tasks is represented. EA is therefore responsible for traveling to the Student's site, for cooperating with the existing PA in order to perform the evaluation, for displaying the questions via a friendly graphical interface to the Student, for allowing the Student to enter his answers, for evaluating the answer and choosing accordingly the next question (adaptive behavior) and finally providing a result of the evaluation.

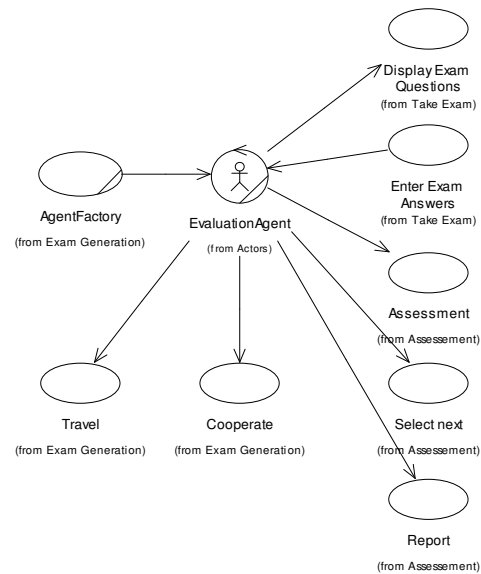

Figure 3: EA and associated tasks

The agent is therefore responsible for performing several concurrent tasks. Each task defines a behavior of the agent. Agent behaviors are represented in our approach as State Chart diagrams (DeLoach et al. 2001). These diagrams contain possible states of the agent and the transitions between states. A state may have a set of associated activities defined as functions (DeLoach 2000), (Sparkman et al. 2001):

result = activity_name(param1, param2, …, paramn)

A transition occurs if the following conditions are true:
- the current state of the task is the initial state of the transition
- the trigger event occurred
- the guard has the logical value *true*
- all the activities of the initial state were performed

The general syntax of a transition is:

Trigger [guard]/ transmission(s)

A transition may generate transmissions. A transmission is either an external message sent to another agent, or an internal event sent to another task of the same agent.

In our approach, each concurrent task is modeled as a state-chart diagram associated to the agent. In Figure 4 an example of a task model for EA is shown.

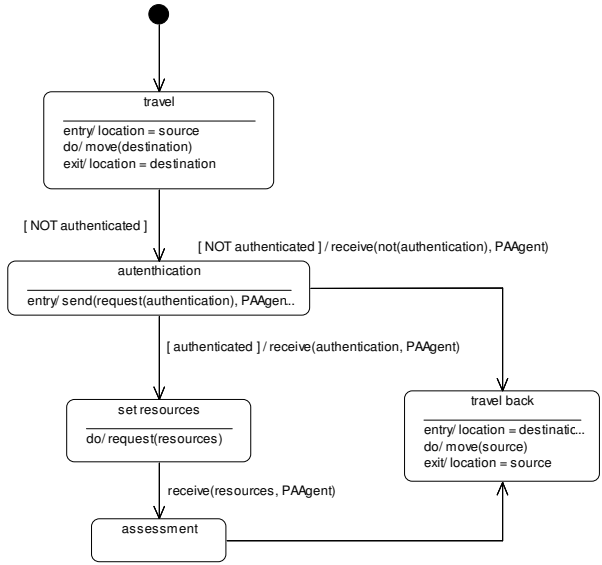

Figure 4: Evaluation Task of EA

**Interaction Model**

Interactions between agents are represented by communication protocols. These protocols are part of the social rules of the organization. The communication protocols details are represented as sequence diagrams (Bauer et al. 2001), (Bergenti and Poggi 2000), (Van Dyke Parunak and Odell 2002). We defined communication protocols between external actors and the system, and also between the agents inside the organization.

**CASE STUDY**

Considering the models developed in the analysis phase, we designed a general architecture of the application. The general architecture was developed as a framework in order to be used by specific applications. The main considered issues were: distribution, reliability, scalability, platform independence, data storage independence, error proof. The architecture has to be therefore well structured and layered.

We considered a multi-layered structure containing well delimited, independent modules. Modules on lower levels provide services to modules on the upper levels.

The main modules of the system are:
- GUI – User Interface Module contaning three submodules:
  o Instructor Interface Module
  o Student Interface Module
  o Admin Interface Module
- BL – Business Logic Module being together with MA (Mobile Agent Module) the core of the system
- MA – Mobile Agent Module
- DAO – Data Access Module – provides primitive data access operations (store, retrieve, update). Isolates the system from the data storage suport assuring independence.
- Utility Server – provides services to other modules. Allows for different configuration settings.

The general architecture is presented in Figure 5.

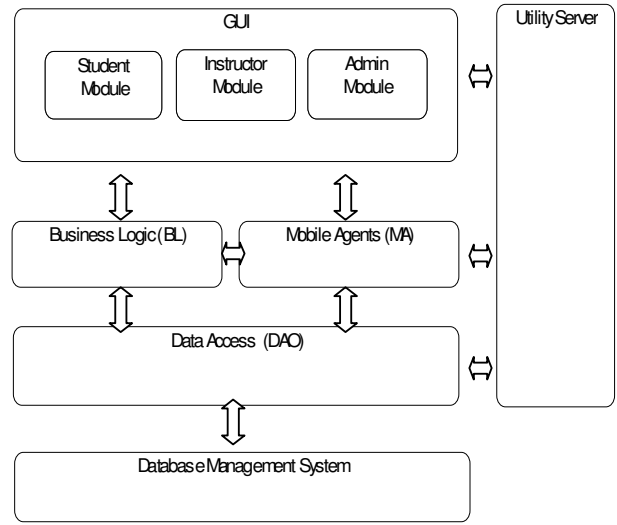

Figure 5: Framework Architecture

The implemented framework uses Java technologies (JSP, Java Beans, etc) and JADE mobile agent platform (JADE 2002). Based on the general framework presented above, a specific Student Assessment Service (SAS) was implemented and integrated in an already existing VLE. In order to define the functionalities of SAS we identified two main functional approaches:
- A Pull (Self-Assessment) Scenario initiated by the Student, who learned a certain section of a specific matter and wants to evaluate his/her knowledge. In this case the Test type is configured by the Student and no record of the assessment is registered in VLE.
- A Push (Exam) Scenario initiated by the Teacher, who enforces a certain Test type for evaluating the students' knowledge level. In this case the configuration is done by the Teacher and the result of the evaluation is recorded in VLE.

We will describe the Self-Assessment module, the Exam module being treated the same way. The Self-assessment module can be further detailed considering sub-modules that can be mapped on some of the social tasks of our system like Assistance, Exam Generation, Taking Exam etc.

For accomplishing the Assistance task we considered the need of a Personal Assistant Agent (PAA) providing an interface for the student to interact with the system. The PAA would be a stationary agent residing on the student's machine. He interacts on the other side with the SAS, communicating the student's requests and providing access to the local resources for the assessment.

We considered on the SAS side the need of a Server Agent (SA) role. The SA should be responsible with managing

PAA's requests, initiating the creation of a specific Evaluation Engine corresponding to the configuration received and also initiating the creation of an Evaluation Agent (EA) responsible with the actual evaluation. SAs are also static agents residing on the SAS's site.

The evaluation has two main phases:
- an offline phase where specific domain knowledge is acquired creating the domain knowledge base. In this phase also the expert answers of the test are analyzed and structured.
- an online phase where the students' answers are analyzed and matched against the expert answers structures.

The off line phase takes place before any assessment is performed.

The need of other components of the SAS is obvious: an Evaluation Engine Factory that should create specific Evaluation Engines for specific assessment configurations. The Evaluation Engine is attached to an EA and provides its ability to analyze the student's answer and to match it against the expert answer, therefore being able to evaluate it.

Another important component is an Agent Factory that actually creates EA's. The EA is a mobile agent, loaded with assessment knowledge (the Evaluation Engine), with a set of questions and expert answers. The EA travels to the student's site and co-operates with the PAA in order to get the assessment done. The EA has an adaptive behavior depending on the student's answers.

We will not focus in this paper on the structure of the Evaluation Engine. The Evaluation Engine will be able to manage different test types like: multiple choice tests, short answers using natural language etc. using a natural language processor based on latent semantic analysis approach.

## CONCLUSION

We are interested in our work to analyze and model specific areas of Virtual Learning Environments and to investigate the most suitable technologies to implement the developed models, particularly mobile agent-based technologies since we are dealing with a distributed and complex environment.

We believe that the methodology we used can be considered a foundation for modeling multi-agent systems. It takes advantage of a goal-driven approach, considers agent-specific issues like roles, tasks and interactions in the analysis phase and can be supported by a well-known modeling language as UML, therefore several off-the–shelf CASE tools being appropriate to be used. We developed a general framework that provided the foundation for a specific Student Assessment Service.

We considered future developments concerning more efficient knowledge representation models for integrating in Evaluation Engines that are suited to be ported by mobile agents.

## REFERENCES

Weiss G. 1999. "Multi-Agent Systems, A Modern Approach to DAI", *MIT Press*.

Bauer B.; J.P.Mueller and J. Odell. 2001. "Agent UML: A Formalism for Specifying Multiagent Interaction", in P. Ciancarini and M. Wooldridge, editors, *Agent-Oriented Software Engineering*. Springer-Verlag Lecture Notes, Berlin, pp.91-103.

Bergenti F., A. Poggi. "Exploiting UML in the Design of Multi-Agent Systems". 2000. In *A. Omicidi, R. Tolksdorf, F. Zambonelli, eds., Engineering Societies in the Agents World - Lecture Notes on Artificial Intelligence*, volume 1972, pp.106-113, Berlin, Germany, Springer Verlag.

DeLoach S.A.; M.F.Wood and C.H.Sparkman 2001, "Multiagent Systems Engineering", *IJSEKE*, Vol.11, No. 3 231-258.

DeLoach S.A. 2000. "Specifying agent Behavior as Concurrent Tasks: Defining the Behavior of Social Agents", *AFIT/EN-TR-00-03*. Technical Report.

Sparkman C.H.; S.A. DeLoach and A.L. Self. 2001 "Automated derivation of Complex Agent Architectures from Analysis Specifications", *AOSE – 2001*, Montreal, Canada.

Van Dyke Parunak H. and J. Odell. 2002. "Representing Social Structures in UML", *Agent-Oriented Software Engineering Workshop II*, Michael Wooldridge, Paolo Ciancarini, and Gerhard Weiss, eds., Springer, Berlin, pp. 1-16.

Wooldridge M., N.R. Jennings and D. Kinny. 2000. „The Gaia Methodology for Agent-Oriented Analysis and Design", *Autonomous Agents and Multi-Agent Systems*, 3(3): 285-312, September.

Zambonelli F., N.R. Jennings, A. Omicini, M. Wooldridge. 2000. „Agent-Oriented Software Engineering for Internet Applications". In *Coordination of Internet Agents: Models, Technologies and Applications*. Springer-Verlag.

Zambonelli F, N.R. Jennings, M. Wooldridge. 2001. „Organisational abstractions for the Analysis and Design of Multi-Agent Systems", in P. Ciancarini and M. Wooldridge, editors, *Agent-Oriented Software Engineering*. Springer-Verlag Lecture Notes in AI Volume 1957, January 2001.

JADE 2002. http://sharon.cselt.it/projects/jade/